\documentclass[11pt,twoside,a4paper]{article}
\usepackage{qmc,psfig}

\newcommand{\lab}[1]{\label{#1}}
\newcommand{\re}[1]{(\ref{#1})}
\newcommand{\bfr}{\begin{flushright}}
\newcommand{\bfl}{\begin{flushleft}}
\newcommand{\efl}{\end{flushleft}}
\newcommand{\efr}{\end{flushright}}
\newcommand{\bc}{\begin{center}}
\newcommand{\ec}{\end{center}}
\newcommand{\be}{\begin{equation}}
\newcommand{\ee}{\end{equation}}
\newcommand{\bea}{\begin{eqnarray}}
\newcommand{\eea}{\end{eqnarray}}
\newcommand{\ba}{\begin{array}}
\newcommand{\ea}{\end{array}}
\newcommand{\edc}{\end{document}}

\newcommand{\ra}{\rightarrow}

\newcommand{\ds}{\displaystyle}
\newcommand{\dsf}{\displaystyle\frac}

\newcommand{\pal}{\partial}

\begin{document}

\title{Complex Diffusion Monte-Carlo method:\\
tests by the simulations of $2D$ electron in magnetic\\ field
and  $2D$ fermions-anyons in  parabolic well}

\author{B. Abdullaev$^1$, M. Musakhanov$^1$, A. Nakamura$^{2}$ }

\instit{
$^1$ Theoretical  Physics Dept, Uzbekistan National University,\\
 Tashkent\\
$^2$ RIISE, Hiroshima University, Japan}

\gdef\theauthor{B. Abdullaev, M. Musakhanov, A. Nakamura }
\gdef\thetitle{Complex Diffusion Monte-Carlo method: tests
...}

\maketitle

\begin{abstract}
We propose a new Complex Diffusion Monte Carlo (CDMC) method for the
simulation of quantum systems with complex  wave function.
In CDMC
the modulus and phase of wave function are simulated both in contrast
to other methods. We successfully test CDMC  by the simulation of the ground
state for $2D$ electron in magnetic field and $2D$ fermions-anyons
in  parabolic well.
\end{abstract}

\section{Introduction}

In the  quantum mechanics, there are several cases where the
 wave function is essentially complex(has imaginary part).
The examples are  $2D$ system of electrons in the external
uniform magnetic field when vector potential has a central-symmetrical
form and system of anyons.
The simulation of these systems is impossible
by using well-known Green
Function Monte Carlo (GFMC) method (see review \cite{ck}) because
GFMC requires the reality of wave function
which is considered as the probability weight during stochastic process.

There have been several attempts \cite{kerbikov,ortiz,lzhang}
to construct a Monte Carlo method for the simulation of quantum systems
with complex wave function. Authors of Ref.\cite{kerbikov} tried to apply
GFMC to this problem and met essential difficulties.
Another method \cite{lzhang} has been developed recently
by using an algorithm without branching that gave the
increasing of the systematical statistical
errors as a function of the simulation time.

The basic difficulty of numerical simulations of fermions is
essentially the same as for the systems with complex wave function.
Their wave function can change the sign and therefore, can not be used
also as the probability weight.
For the simulation of fermionic systems a widely used
fixed node Monte Carlo method has been developed
\cite{anderson}  (see also \cite{renolds}). Recently the
constrained path Monte Carlo method was proposed\cite{szhang}.
In both methods one assumes the restriction on the random walks
connected with the uncertainty in the space localization of the wave
function node surface.
CDMC method is very close to the fixed
phase diffusion Monte Carlo method \cite{ortiz}  which was
applied to $2D$ electrons in magnetic field.
In the framework of diffusion Monte Carlo  method
one simulates only  modulus of the system  wave function. The phase of
this wave function is not simulated  but treated to be
fixed and equal to the  phase of Laughlin's wave function \cite{l}.
Here we propose a new CDMC method
{\em including also the simulation of the phase of the system wave function}
\cite{abdullaev1,abdullaev2}.
In the $2D$ space we have unique tool --  so-called
anyons, particles (bosons or fermions) with additional
gauge interaction which provide required  statistical properties \cite{wu}.
So, by tuning the coupling constant we may arrive to the fermions
starting from the bosons.
It looks very attractive to apply the anyons for the simulation of
$2D$ fermions because statistical vector potential
 that gives anyonic property is relatively smooth and continuous
\cite{wu,laughlin}.
The remaining main problem is that the wave function of anyonic system
contains an imaginary part.

The first test of CDMC
reproduces the ground state energy and its degeneracy
on the orbital quantum number $m$ of electron in magnetic field.
Also the simulated wave function
is in exact correspondence with analytical prediction (Fig.\re{fig-1,2}).
As the second
test of CDMC, we simulate ground state energy for $2D$ fermions
and anyons in parabolic well.
The system of $2D$ fermions has a well-known analytical
solution for this energy as a function of the number of fermions.
The simulation is done for the fermionic systems with the number of particles
from two to ten. We have found a good agreement of the numerical results
with analytical ones for the numbers of the particles great than four
(Tables \re{tab-1}, \re{tab-2}).
This observation of simulation allows us to hope that
the CDMC is a good tool for the simulation of $2D$ fermion systems with big
number of particles.
The simulated ground state energy for anyons (Table \re{tab-3})
we compare with variational
calculation \cite{abdullaev3}.

\section{ Ground state of $2D$ electron in magnetic field }

Hamiltonian of the electron with  $\vec A=[\vec H\vec r]/2$
(in the gauge
$\vec p\vec A -\vec A\vec p=div \vec A=0$)
has a form:
\be
\hat H=\dsf{\vec p\ ^2}{2M}+\dsf{|e|}{Mc}\vec A\vec p+
\dsf{|e|^2}{Mc^2}\vec A^2 .
\lab{2}
\ee
Here  $\vec p=-i\hbar \vec
\nabla$,  $\vec \nabla= ({\pal}/{\pal x}, {\pal}/{\pal y})$.

Let us introduce a complex distribution function
\be
f(\vec r,t)=\Psi_T^*(\vec r)\Psi(\vec r,t).
\lab{3}
\ee
Here $\Psi_T^*(\vec r)$ is
a complex conjugated trial wave function  of electron in the magnetic field.
Schr\"odinger equation for $\Psi(\vec r,t)$ with
imaginary time ( expressed in $\hbar$ units ) is:
\be
-\dsf{\pal \Psi(\vec r,t)}{\pal t}=(\hat H-E_T)\Psi(\vec r,t).
\lab{4}
\ee

The equation for the
distribution function $f\equiv f(\vec r,t)$ is:
\bea
-\dsf{\pal f}{\pal t}&=& -D\Delta f+D\vec \nabla(f\ Re \vec F_Q(\vec r))+
i[\vec \nabla(D f\ Im \vec F_Q(\vec r))-
\nonumber
\\
&& -\dsf{\hbar|e|}{Mc}\vec A\vec \nabla f]+
(E_L(\vec r)-E_T)f.
\lab{12}
\eea
Here we have
\be
\vec F_Q(\vec r)=2\Psi_T^{*-1}(\vec r)\vec\nabla\Psi_T^{*}(\vec r),
\lab{9}
\ee
\be
E_L(\vec r)=\Psi_T^{*-1}(\vec r)\hat H'\Psi_T^{*}(\vec r),
\lab{7}
\ee
\be
\hat H'=\dsf{1}{2M}(\vec p-\dsf{|e|}{c}\vec A)^2
\lab{8}
\ee
where we putted
$\vec p=-i\hbar\vec \nabla$,
$D=\dsf{\hbar^2}{2M}$, $\Delta=\vec \nabla^2$ and took into account
\be
\vec F_Q(\vec r)=Re \vec F_Q(\vec r)+i Im \vec F_Q(\vec r)
\lab{11}
\ee
when  $\Psi_T^*(\vec r)$ is complex.

Following  \cite{renolds},
in the limit $\tau \ra 0$  for the time step
of equation \re{12}
we assume $\vec F_Q(\vec r)\equiv \vec F_Q(\vec
r\ ')$  where $\vec r$ corresponds to time point
$t+\tau$ and $\vec r\ '$ to  $t$.

Let us introduce a new quantity
\be
\vec A_Q(\vec r,\vec r\ ')=\dsf{1}{2}Im \vec F_Q(\vec r\ ')-
\dsf{\hbar|e|}{2DMc} \vec A (\vec r).
\lab{13}
\ee
Then at $\tau \ra 0$ the Green function of equation \re{12} has a form:
\bea
G(\vec r,\vec r\ ';\tau)=G_{1}(\vec r,\vec r\ ';\tau)
\exp\left[-\tau(E_L(\vec r)-E_T)\right] \times \nonumber\\
\times \exp\left[i\vec A_Q(\vec r,\vec r\ ')
(\vec r-\vec r\ '-D\tau Re \vec F_Q(\vec r\ '))\right]
\lab{14}
\eea
where
\be
G_{1}(\vec r,\vec r\ ';\tau)=\dsf{\exp[D\tau\vec A_Q^2(\vec r,\vec r\ ')]}{
4\pi D\tau}
\exp\left[-\dsf{(\vec r-\vec r\ '-D\tau Re \vec F_Q(\vec r\ '))^2}{
4D\tau}\right].
\lab{14a}
\ee
The Green function $G(\vec r,\vec r\ ';\tau)$ \re{14}
and distribution function $f$  \re{3} are the complex functions and
related by usual integral equation:
\be
f(\vec r,t+\tau)=\int d\vec r\ ' G(\vec r,\vec r\ ';\tau)f(\vec r\ ',t).
\lab{15}
\ee
From \re{15} the modulus and  phase of $f$ at $t+\tau$ are determined by
the modulus and phase of Green function and of ones of $f$ at $t$.

If $\Psi_T^*(\vec r)$ is complex then the energy $E_L(\vec r)$ is a complex
too.
Therefore, the last two exponents in \re{14} have a form:
\bea
&&\exp\left[-\tau(Re E_L(\vec r)-E_T)\right]
\nonumber\\
&&\times\exp\left[i\vec A_Q(\vec r,\vec r\ ')
(\vec r-\vec r\ '-D\tau Re \vec F_Q(\vec r\ '))-i\tau Im E_L(\vec r)\right].
\lab{16}
\eea
CDMC algorithm \cite{abdullaev1} is:
\\
1. We prepare $N_c$ initial configurations $\vec r$
in which the electron has the random
and uniform distribution in a position.
\\
2. We perform the quantum drift and diffusion of the particle:
 $
\vec r_k=\vec r_k\ '+D\tau Re \vec F_Q(\vec r_k\ ')+\chi,
$
for example, from $k$ -th configuration.
Here  $\chi$ is Gaussian random number having zero mean value and
the dispersion $2\sqrt{D\tau}$.
\\
3. The transition into new space point is accepted
with probability
$
P(\vec r\ '\ra \vec r, \tau)\equiv \min (1,W(\vec r, \vec r\ '))
$
where
$
W(\vec r, \vec r\ ')=\dsf{|\Psi_T(\vec r)|^2G_1(\vec r\ ', \vec r; \tau)}{
|\Psi_T(\vec r\ ')|^2G_1(\vec r, \vec r\ '; \tau)}.
$
$G_1(\vec r, \vec r\ '; \tau)$
is given by \re{14a}.
If the transition of electron is accepted then in accordance with \re{16}
it has a new phase; if not  then the electron keeps its old phase.
\\
4. After changing the electron position from $k$-configuration
into new space point, one calculate  $Re E_L(\vec r_k)$, $Im
E_L(\vec r_k)$ and other quantities for measurements.
\\
5. By using the first exponential factor in \re{16} we calculate the
multiplicity $M_k$
according to the formula
$
M_k=\exp[-\tau(Re E_L(\vec r_k)-E_T)].
$
If $M_k$ is not integer  we add an uniformly distributed random number
between 0 and 1 to it and take $M_k$ equal to nearest integer.
\\
6. If $M_k\ne 0$  then $M_k$ copies of new $k$-th configuration
are put in the list of the new $N$ configurations which becomes
the initial one
at the next step $\tau$ in the integration of the diffusion equation.
If $M_k=0$  then there is no $k$-th  configuration in the list of new
$N$ configurations for the next time step $\tau$.
All quantities of interest such
as $Re E_L(\vec r_k)$, $Im E_L(\vec r_k)$ etc.
are multiplied by the factor $M_k$ when we calculate their mean values.
\\
7. We repeat the steps (2) - (6) until all
$N_c$ configurations are not overlooked and electrons on these
configurations are not simulated on the displacement.
\\
8. We calculate a mean energy and
other mean quantities on the $N$ configurations obtained at
the step (6)  at the time step  $\tau$
in accordance with formula \re{23}.
\\
9. We repeat the points (1) - (8) of algorithm an integer number of time
steps $\tau$ of integration of diffusion equation.
After that one determine the mean values of quantities
$Re \overline E$, $Im \overline E$ and others on this integer number too.
New value of $E_T$ is redetermined in accordance with the formula
$(E_T)_{new}=[(E_T)_{old}+Re \overline E]/2$ with the assumption
$Im \overline E <<Re \overline E$.
The integer number of time steps $\tau$ represents one time block
$\Delta t$.
\\
10. Every time block  $\Delta t$ has the $N_c$ initial configurations.
The list of these configurations is filled  by randomly chosen configurations
from just ended time block. The random choice of configurations consists
from two steps:\\
a) a random choice of configuration in the every time step $\tau$
in the time block $\Delta t$;\\
b) in the case $N_c$ larger than number of the time steps $\tau$ in the
time block  $\Delta t$,
 other configurations are taken randomly from the last time step $\tau .$
\\
In this manner  filled list of the $N_c$ configurations will be initial one
for the next time block $\Delta t$.
\\
11. The big number repeating of time blocks $\Delta t$  essentially
decreases  a correlation between configurations in neighbor
time blocks and provides a right calculation of mean quantities.

The calculation of the mean quantities by complex distribution function
$f(\vec R,t)$ at the time $t$ of the running process is occurred
by the formula \cite{kerbikov}:
\bea
<F(t)>&=&\dsf{\ds\sum_{i=1}^M\exp[i\alpha(\vec R_i(t)]F(\vec R_i(t))}{
\ds\sum_{i=1}^M\exp[i\alpha(\vec R_i(t)]}
\lab{21}\\
&\approx &\dsf{\ds\sum_{i=1}^Me^{i\alpha_G(\vec R_i(t), \vec R_j(t-\tau))}
F(\vec R_i(t))}{
\ds\sum_{i=1}^Me^{i\alpha_G(\vec R_i(t), \vec R_j(t-\tau))}}.
\lab{23}
\eea
Here $\vec R_i(t)$ is  $\vec r_1,\vec r_2,...,\vec r_N$,
$N$ is the number of particles
and
$M$ is the number of configurations at time point $t$.
The particle coordinates $\vec r_1,\vec r_2,...,\vec r_N$
in \re{21}, \re{23} are weighted with probability $|f(\vec R_i,t)|$
and  $\alpha(\vec R_i(t))$ is a phase of the distribution
function $f(\vec R_i,t)$and $\alpha_G$ is
the phase of  Green function \re{16}.

The step from Eq. \re{21} to Eq. \re{23} is rather nontrivial \cite{fantoni}
and can be proved with account the properties of
the phase of  Green function \re{16}
$\alpha_G$. Because at $\tau\ra 0$
it  has an asymptotic $\alpha_G\sim\tau^{1/2}$ (see \cite{abdullaev1})
and oscillates randomly around zero.
In the following, the comparison of the results of
numerical calculation of ground state energy of $2D$ fermions
by the formula \re{21} (Table \re{tab-2})  and by the formula  \re{23}
(Table \re{tab-1}) confirms this suggestion.
\begin{figure}[tb]
\centerline{\psfig{file=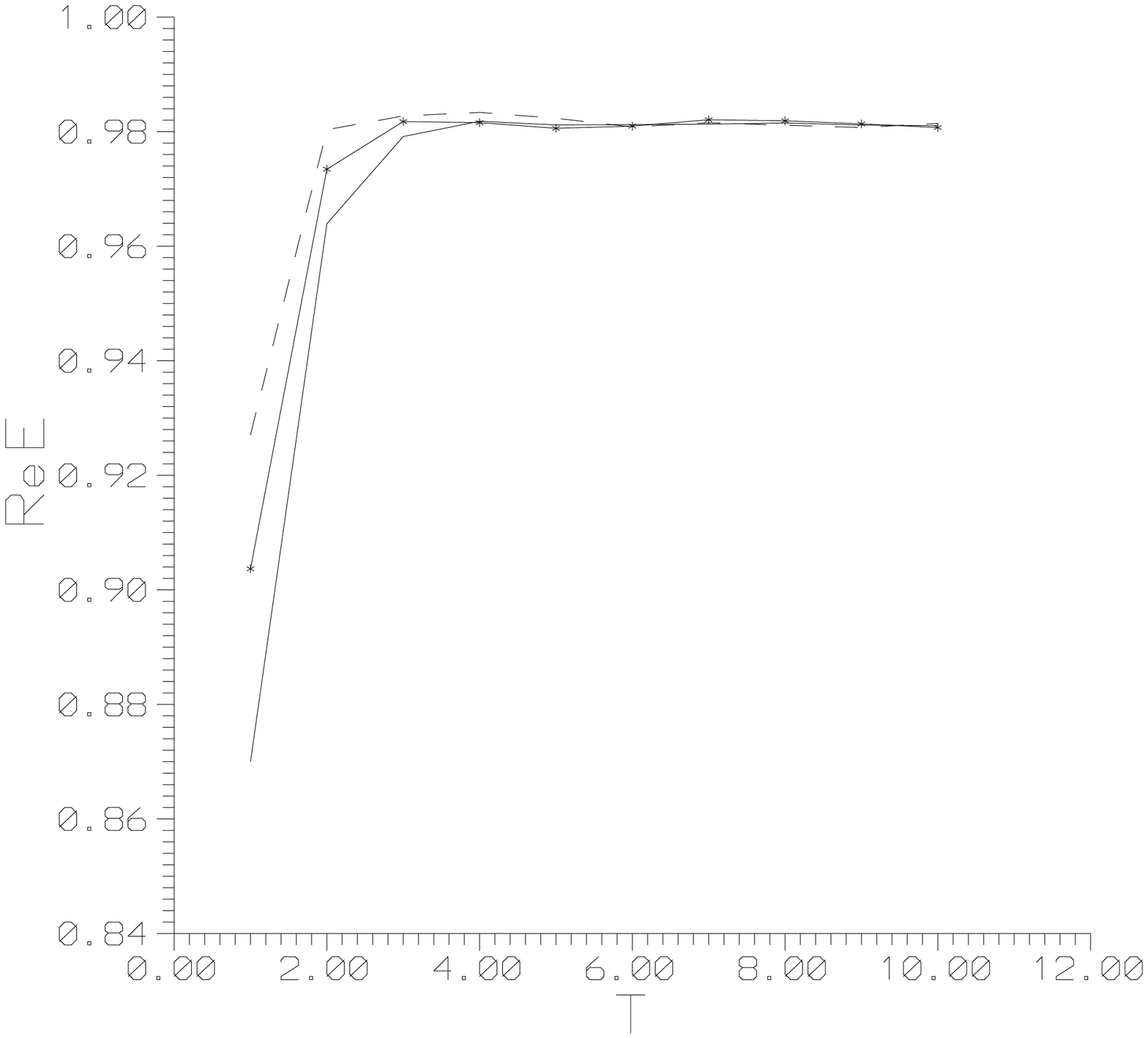,width=55mm}
\psfig{file=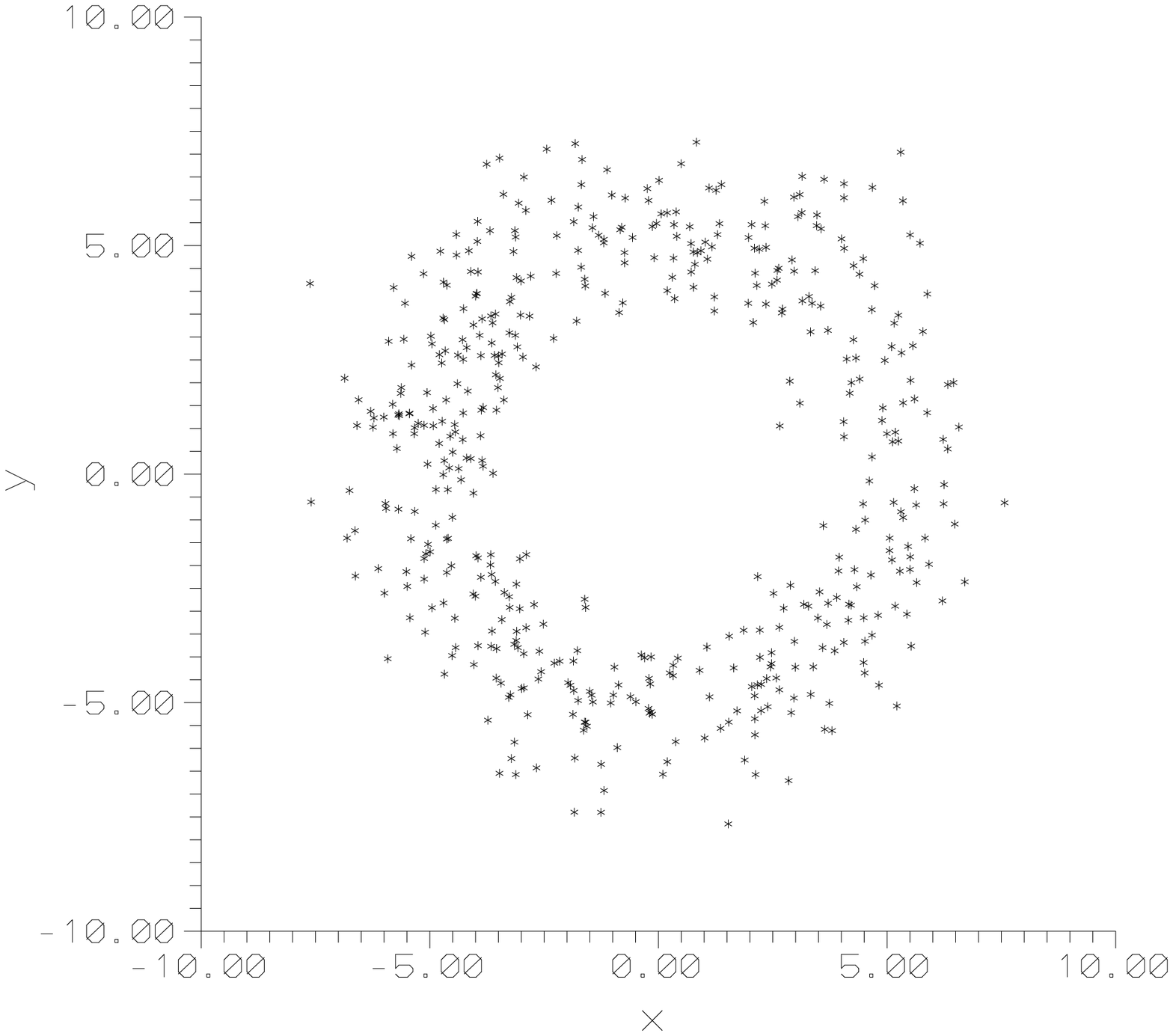,width=55mm}}
\caption{Left panel:
The real part
of the ground state energy $Re \overline E$
as function of the
number $T$ of time blocks $\Delta t$
for different orbital quantum numbers $m$.
Solid line, marked solid line and dashed line
correspond $m=0$, $m=4$ and $m=8$, respectively.
Here $\alpha = \beta =\gamma = 1.0$.
Right Panel:
The simulation result for
ground state spatial distribution of
the electron in uniform magnetic field  at $m=13$. The  length
unity is $a_c$, initial number of points $N_c=500$
and parameters $\alpha = \beta =\gamma =1$.}
\label{fig-1,2}
\end{figure}

The trial wave function for the ground state of $2D$ electron in
magnetic field is:
\be
\Psi_T^*(\vec r)=\dsf{C}{a_c}\left(\dsf{\alpha
x+i\beta y}{a_c}\right)^m \exp\left(-\gamma\dsf{(x^2+ y^2)}{4a_c}\right).
\lab{26}
\ee
$C$ - normalizing constant, $a_c=(\hbar/Mw_c)^{1/2}$ - magnetic length
and $w_c=|e|H/Mc$ -
 cyclotron frequency, $x$ and  $y$ - components of
electron coordinate,
$m$ is the orbital quantum number, and $\alpha$, $\beta$ and $\gamma$
are arbitrary numerical constants.
When $\alpha=\beta=\gamma=1$ \re{26} is an exact
ground state wave function.
We take $a_c$, $\hbar w_c/2$, and $2/\hbar w_c$
as length, energy and time units, respectively.
A number of time blocks $\Delta t$  is ten.
For every  $\Delta t$  the initial number of configurations $N_c=1000$
and number of time steps $\tau$ is
equal 100. Everywhere  $\tau$ was equal 0.01.

Left panel of Fig.\re{fig-1,2} presents
$Re \overline E$ as functions of number time blocks
$T$. Right panel of Fig.\re{fig-1,2} is the ground state  spatial distribution
of electron after simulation for $m=13$.
From this Fig. we see good correspondence with
exact solution \cite{galicki} for the spatial
distribution which is a ring  with
mean radius  $(2m+1)^{1/2}$ and wide 1 in $a_c$ magnetic length units.

\section{ Ground state for
$2D$ fermions and anyons in  parabolic well}

Hamiltonian of anyons in the $2D$ parabolic well has a form:
\be
\hat H=\dsf{1}{2M}\ds\sum_{i=1}^N(\vec p_i+\vec A(\vec r_i))^2+
\dsf{M\omega_o^2}{2}\ds\sum_{i=1}^N\vec r_i^2.
\lab{f1}
\ee
Here $M$ is the mass of particle,  $\vec p_i=-i\hbar \vec
\nabla _i$  where
$\vec \nabla _i=({\pal}/{\pal x_i}, {\pal}/{\pal y_i})$,
$\omega_o$ is a characteristic frequency of free
particles in  parabolic well and $\vec r_i$ is radius vector of
$i$-th particle, $N$ is number of particles.

Vector potential for anyons $\vec A(\vec r_i)$ \cite{wu,laughlin}
in \re{f1} is
\be
\vec A(\vec r_i)=\hbar\nu\ds\sum_{j>i}^N\dsf{\vec z \times\vec r_{ij}}
{|\vec r_{ij}|^2}.
\lab{f2}\ee
Here $\vec z$ is the unit vector perpendicular to $2D$ plane and $\nu$ -
anyonic fractional statistics factor (spin of the anyon).
In the bosonic representation of anyons
$\nu=0$ gives noninteracting bosons and $\nu=1$ gives
 free fermions.

We introduce
 a complex distribution function
\be
f(\vec R,t)=\Psi_T^*(\vec R)\Psi(\vec R,t)
\lab{f3}\ee
where $\vec R$ stands for the coordinates of all particles
and the wave function
$\Psi(\vec R,t)$ satisfies a Schr\"{o}dinger equation \re{4} with
hamiltonian \re{f1}.
\noindent

  For a bosonic representation of anyons we take
(conjugated) trial wave function in the form:
\be
\Psi_T^*(\vec R)=\prod_{i=1}^N\Psi_T^*(\vec r_i).
\lab{f5}\ee

  The distribution function $f\equiv f(\vec R,t)$ satisfies a diffusion
equation:
\bea
-\dsf{\pal f}{\pal t}= -D\sum_{i=1}^N\Delta_if+D\sum_{i=1}^N\vec\nabla_i
(f \mathrm{Re} \vec F_Q(\vec r_i))+
\nonumber \\
 +i\ds\sum_{i=1}^N[\vec\nabla_i(Df \mathrm{Im} \vec F_Q(\vec r_i))-
\dsf{\hbar}{M}\vec A(\vec r_i)\vec \nabla_i f]+(E_L(\vec R)-E_T)f.
\lab{f6}
\eea
Here $D={\hbar^2}/{2M}$ and $\Delta_i=\vec\nabla_i^2$.
When a time step integration of diffusion equation \re{f6}
$\tau$ goes to zero, the Green function for this equation
has a form:
\bea
G(\vec R,\vec R\ ';\tau)=\dsf{\exp[D\tau\ds\sum_{i=1}^N
\vec A_Q^2(\vec r_i,\vec r_i\ ')]}{(4\pi D\tau)^N}\times
\nonumber \\
\times\exp\left[-\dsf{\ds\sum_{i=1}^N(\vec r_i-\vec r_i\ '-D\tau
\mathrm{Re} \vec F_Q(\vec r_i\ '))^2}{
4D\tau}\right]\exp\left[-\tau(E_L(\vec R)-E_T)\right]\times
\nonumber \\
\times
\exp\left[i\ds\sum_{i=1}^N\vec A_Q(\vec r_i,\vec r_i\ ')
(\vec r_i-\vec r_i\ '-D\tau \mathrm{Re} \vec F_Q(\vec r_i\ '))\right].
\lab{f7}
\eea
In the expressions \re{f6} and \re{f7}
$\vec F_Q(\vec r_i)$  determines by formula \re{9} where radius vector
and operator nabla have index $i$, the energy $E_L(\vec R)$ determines by
formula \re{7} with wave function  $\Psi_T^{*}(\vec R)$ and hamiltonian
$\hat H'$ (the hamiltonian \re{f1} with changing of
sign in circle brackets (see \re{8} for determination of $\hat H'$)).
$\vec F_Q(\vec r_i)$ for the complex wave function  $\Psi_T^*(\vec r_i)$
has the real and imaginary parts.

In \re{f7} we have introduced a new quantity
\be
\vec A_Q(\vec r_i,\vec r_i\ ')=\dsf{1}{2}\mathrm{Im} \vec F_Q(\vec r_i\ ')-
\dsf{1}{\hbar}\vec A (\vec r_i).
\lab{f12}\ee
In the expressions \re{f7} and \re{f12}
vectors $\vec R$ and  $\vec r_i$ correspond to the time point
$t+\tau$ and vectors $\vec R\ '$,  $\vec r\ '$ to the time point $t$.

As the energy $E_L(\vec R)$ is complex, so
the expression for two last exponents in \re{f7} has a form:
\bea
\exp\left[-\tau(\mathrm{Re}E_L(\vec R)-E_T)\right]\times \nonumber\\
\times
\exp\left[i\ds\sum_{i=1}^N\vec A_Q(\vec r_i,\vec r_i\ ')
(\vec r_i-\vec r_i\ '-D\tau \mathrm{Re} \vec F_Q(\vec r_i\ '))-
i\tau\mathrm{Im}E_L(\vec R)\right].
\lab{a12}
\eea
The ground state trial wave function for one
anyon in parabolic well and in the mean magnetic field \cite{fetter}
with vector potential
\be
\overline {\vec{A}}(\vec r)=\rho\pi\hbar\nu(\vec z\times\vec r) =
\dsf{1}{2}\vec B\times\vec r
\lab{f13}\ee
which is generated by the average density of particles $\rho$
can be chosen as:
\be
\Psi_T^*(\vec r_i)=C\exp\left(-\alpha\dsf{(x_i^2+y_i^2)}{2R_o^2}\right)
\exp\left(-\dsf{(x_i^2+ y_i^2)}{4a_H^2}\right).
\lab{f14}
\ee
Here $\vec B=2\pi\rho\hbar\nu\vec z$ is fictitious uniform mean magnetic
field. Constant C is for normalization and $R_o=(\hbar/M\omega_o)^{1/2}$.
Density $\rho$ determines by formula
$\rho=1/\pi r_o^2$  where $r_o$ is the mean distance between particles. We
assume (see \cite{abdullaev2}) that $r_o$ is equal to $R_o$.
We express the energies in terms of $\hbar\omega_o$ and the lengths in
terms of $r_o$.
\begin{figure}[tb]
\centerline{\psfig{file=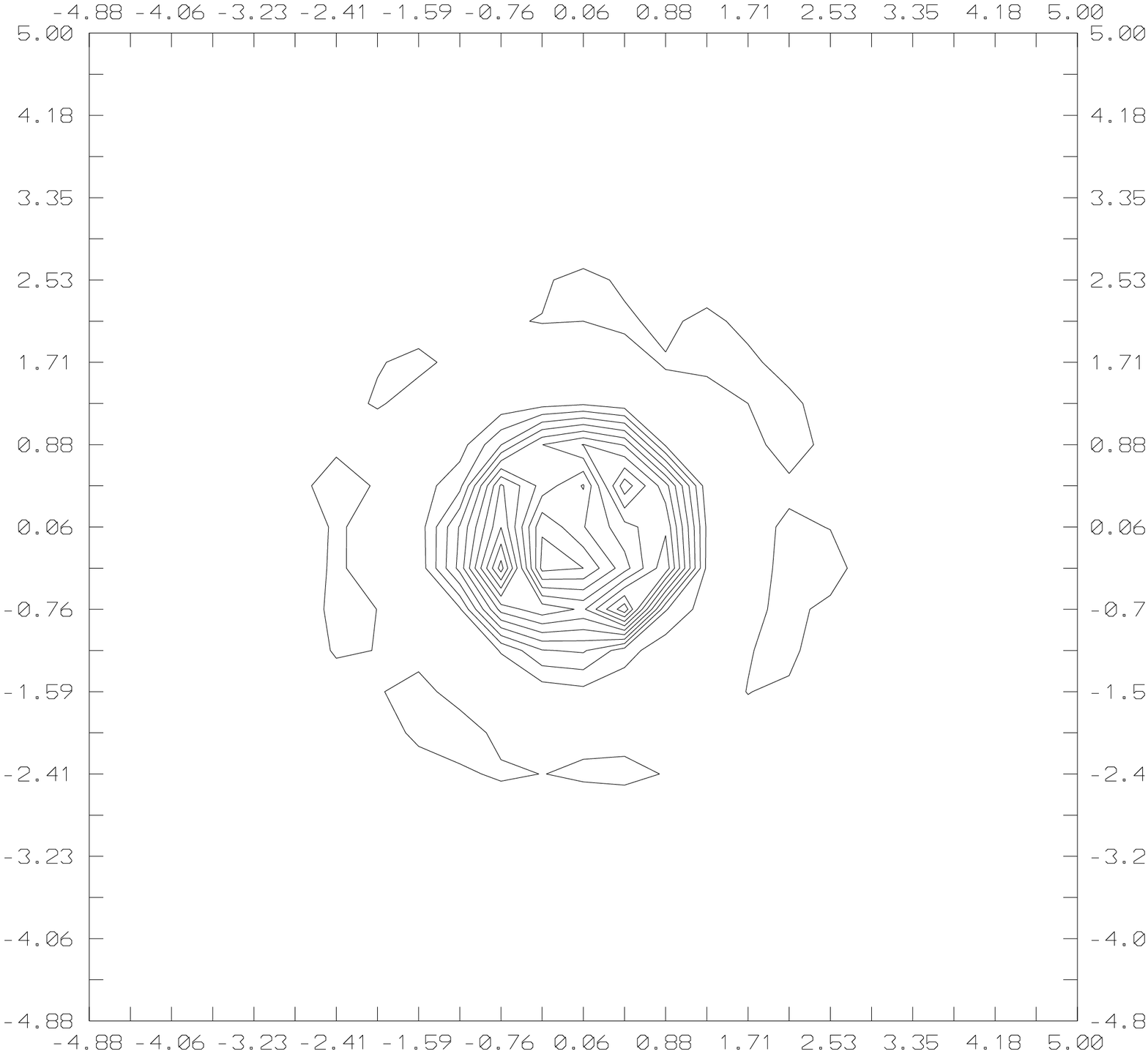,width=55mm}
\psfig{file=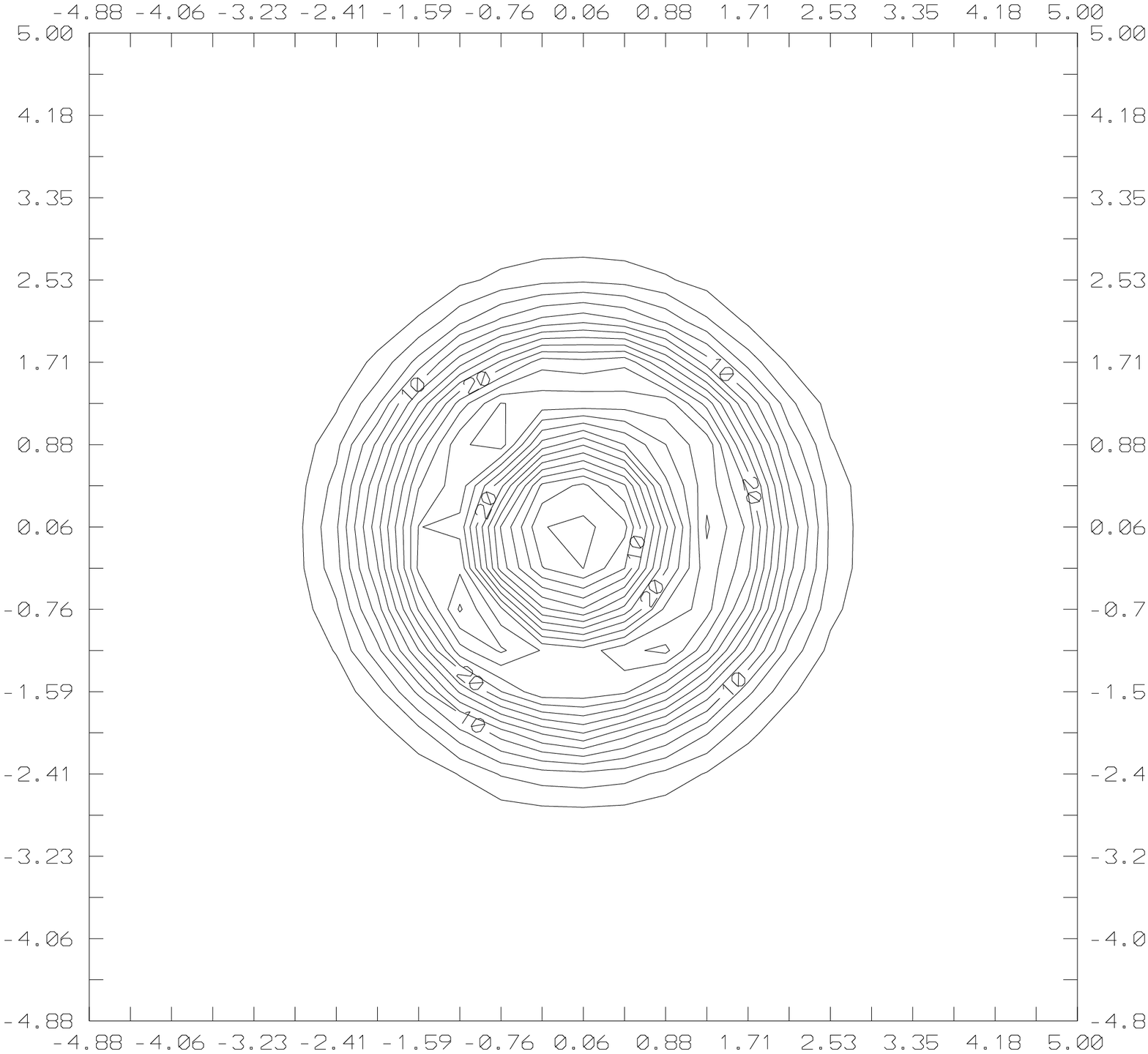,width=55mm}}
\caption{
Square of module of one particle
ground state wave function  for 6 fermions
$|\Psi(\vec r)|^2=\int|\Psi(\vec r=\vec r_1,\vec r_2,\ldots,\vec r_6)|^2
d\vec r_2 d\vec r_3 \ldots  d\vec r_6$.
Left panel: result of the simulation.
 Right Panel: calculations with Slater determinant.  }
\label{fig-3,4}
\end{figure}
We calculate anyons ground state energy at $\nu=0.2,0.4,0.6,0.8$.
The total simulation time $T$ is equal 40 time blocks $\Delta t$.
Anyons become fermions at $\nu=1$.

For every  $\Delta t$ $N_c=1000$ and the number of time steps $\tau$ is 30.

We take here optimal time step $\tau_0$ in accordance with \cite{abdullaev2}.
We find that the imaginary parts of the simulated mean quantities
for all numbers fermions and anyons are
essentially less than their real parts.

The results of the simulation for $2D$ fermions-anyons are given in tables
\re{tab-1} - \re{tab-3}. In tables \re{tab-1} and \re{tab-2} exact
energies $E$ were calculated
by occupying the available lowest states of $2D$ parabolic well
while the minimization of
the simulated energy over
parameters $\alpha$ gives the $E_{SIM}$.
The comparison of the square of module of
simulated one particle ground state wave function
for 6 fermions  in parabolic well with the square of the same module
calculated by Slater determinant is given in the figures \re{fig-3,4}.
From tables \re{tab-1} and \re{tab-2} we see that for four and
bigger numbers of $2D$ fermions we have good simulation results.
\begin{table}[tb]
\begin {center}
\begin{tabular}{|c|c|c|c|c|c|c|} \hline
    N      & $E$ & $E_{SIM}$ & $\overline{r}$ & $\overline{r^2}$ &
$\alpha$   &$\tau_0$ \\ \hline
    2      &        3    &  1.842                &       0.935    &   0.8852
& -0.593 &0.08 \\
           &             & $\pm1.297\cdot10^{-3}$ & $\pm6.994\cdot10^{-4}$
& $\pm1.319\cdot10^{-3}$  &             &\\ \hline
    4      &        8    & 7.399                 & 1.236
& 1.919    &- 0.3        &0.04 \\
   {}      &        {}   &{$\pm4.092\cdot10^{-2}$}&{$\pm1.844\cdot10^{-2}$}
& {$\pm4.562\cdot10^{-2}$} &  & \\ \hline
  {6}      &    {14}     &   {14.492}            &    {1.619}
& {3.275}  &  {-0.092}    &0.02 \\
   {}      &     {}      &{$\pm3.136\cdot10^{-2}$}&{$\pm4.261\cdot10^{-3}$}
&{$\pm1.564\cdot10^{-2}$} &    {}                 & \\ \hline
    {8}    &    {22}     &        {22.383}       &   {2.020}
& {5.112}  & {-0.046}     &0.015 \\
   {}      &    {}       &{$\pm8.569\cdot10^{-2}$}&{$\pm5.821\cdot10^{-3}$}
&{$\pm2.790\cdot10^{-2}$} &        {}             & \\ \hline
    {10}   &     {30}    &        {30.905}       & {2.601}
&  {8.599} &    {-0.084}   & 0.005\\
     {}    &    {}       &{$\pm2.010\cdot10^{-1}$}&{$\pm2.839\cdot10^{-2}$}
&{$\pm2.095\cdot10^{-1}$} &          {}           & \\ \hline
\end{tabular}
\end{center}
\vskip -.5cm
\caption{The ground state energies for the fermions in $2D$ parabolic well.
Here
$N$ - numbers of particles, $E$ - analytically calculated  ground
state energies,
$E_{SIM}$ - results of the simulation calculated by formula \protect \re{23}
(both in $\hbar\omega_o$ units),
$\overline r$ - the  simulated mean
radius (in $r_o$ units), $\overline {r^2}$ - the
simulated mean square radius
(in $r_o^2$ units), $\alpha$ - the numerical parameters in the  wave
function \protect \re{f14} that give a minimum $E_{SIM}$
and $\tau_0$ - the optimal time steps. All
simulated quantities and their deviations from mean values were averaged
over 30 last time blocks $\Delta t$ when $E_{SIM}$ and mean population
number $N_p$ have had
relatively stable values ($N_p\approx 1000$). }
\vskip -.5cm
\lab{tab-1}
\end{table}
Good results for the simulation of ground state for the electron in magnetic
field(see figures \re{fig-1,2}) and for the systems of $2D$ fermions
in parabolic well (tables \re{tab-1}, \re{tab-2}, figures \re{fig-3,4})
(for the number  of
particles bigger than four) allows
us to hope that CDMC provides
the successful simulation of  $2D$ fermionic systems with Coulomb interaction.

We are grateful to S.Fantoni, M.-P. Lombardo, G.Ortiz and G.Senatore
for useful discussions.
One of us (B.A.) would like to thank
organizers  for the support of his participation in  Quantum Monte Carlo
2001 workshop in Trento.

\newpage
\begin{table}[tb]
\begin {center}
\begin{tabular}{|c|c|c|c|c|c|c|c|}\hline

    N      & $E$ & $E_{SIM}$ & $\overline{r}$ & $\overline{r^2}$ &
$\alpha$   &$\tau_0$ \\ \hline
    2      &        3    &  1.842                &       0.935    &   0.8847
& -0.593 &0.08 \\
           &             & $\pm1.192\cdot10^{-3}$ & $\pm6.974\cdot10^{-4}$
& $\pm1.418\cdot10^{-3}$  &             &\\ \hline
    4      &        8    & 7.402                 & 1.242
& 1.936    &- 0.3        &0.04 \\
   {}      &        {}   &{$\pm2.374\cdot10^{-2}$}&{$\pm1.202\cdot10^{-2}$}
& {$\pm3.149\cdot10^{-2}$} &  & \\ \hline
  {6}      &    {14}     &   {14.493}            &    {1.612}
& {3.245}  &  {-0.092}    &0.02 \\
   {}      &     {}      &{$\pm4.160\cdot10^{-2}$}&{$\pm5.206\cdot10^{-3}$}
&{$\pm2.219\cdot10^{-2}$} &    {}                 & \\ \hline
    {8}    &    {22}     &        {22.358}       &   {2.011}
& {5.075}  & {-0.046}     &0.015 \\
   {}      &    {}       &{$\pm6.639\cdot10^{-2}$}&{$\pm6.379\cdot10^{-3}$}
&{$\pm3.065\cdot10^{-2}$} &        {}             & \\ \hline
    {10}   &     {30}    &        {30.896}       & {2.598}
&  {8.574} &    {-0.084}   & 0.005\\
     {}    &    {}       &{$\pm2.303\cdot10^{-1}$}&{$\pm4.129\cdot10^{-2}$}
&{$\pm2.941\cdot10^{-1}$} &          {}           & \\ \hline
\end{tabular}
\end{center}
\vskip -.5cm
\caption{The ground state energies for the fermions in $2D$ parabolic well.
Here
$E_{SIM}$ - results of the simulation calculated by formula \protect \re{21},
the determinations of other quantities are the same as in table
\protect \re{tab-1} and $N_p\approx 1000$.}
\vskip -.5cm
\lab{tab-2}
\end{table}

\newpage

\begin{table}[tb]
\begin {center}
\begin{tabular}{|c|l|c|c|c|c|c|c|c|} \hline
    $\nu$  & $E_v$ & $E_{SIM}$ & $\overline{r}$ & $\overline{r^2}$ &
$\alpha$   &  $\tau_0$ \\ \hline
 0.2 & 32.00 & 30.815 & 1.353 & 2.282 & 0.500 & 0.001 \\
     & & $\pm3.222$ & $\pm7.078\cdot10^{-3}$ & $\pm2.430\cdot10^{-2}$ &
& \\ \hline
 0.4 & 42.33 & 40.525 & 2.081 & 5.508 & 0.500 & 0.001 \\
    & & $\pm0.268$ & $\pm3.611\cdot10^{-2}$ & $\pm2.040\cdot10^{-1}$
&& \\ \hline
 0.6 & 50.59 & 48.371 & 2.260 & 6.185 & 0.500 & 0.0075 \\
     & & $\pm1.605$ & $\pm5.124\cdot10^{-2}$ & $\pm2.771\cdot10^{-1}$ &
& \\ \hline
 0.8 & 57.68 & 54.119 & 3.085 & 11.984 & 0.177 & 0.005 \\
     & & $\pm1.142$ & $\pm7.307\cdot10^{-2}$ & $\pm6.544\cdot10^{-1}$&
& \\ \hline
 1.0 & 64.00 & 65.401 & 3.646 & 16.872 & -0.050 & 0.005 \\
     & & $\pm0.914$ & $\pm6.379\cdot10^{-2}$ &
$\pm7.028\cdot10^{-1}$ & & \\ \hline
\end{tabular}
\end{center}
\vskip -.5cm
\caption{The ground state energy as function of anyonic factor $\nu$
for 16 anyons in $2D$ parabolic well.
Here
$\nu$ - anyonic factor, $E_v$ - a variational ground
state energies \protect \cite{abdullaev3}
(in $\hbar\omega_o$ units) and
$N_p\approx 1000$ (except $N_p\approx 2000$ at
$\nu=0.6$). The determinations of other quantities are the same as in tables
\protect \re{tab-1}  and  \protect \re{tab-2}.
 }
\vskip -.5cm
\lab{tab-3}
\end{table}

\end{document}